# Interactional processes for stabilizing conceptual coherences in physics

Brian W. Frank, Department of Physics, Middle Tennessee State University, Murfreesboro, TN 37132

Rachel E. Scherr, Department of Physics, Seattle Pacific University, Seattle, WA 98119

*Abstract*. Research in student knowledge and learning of science has typically focused on explaining conceptual change. Recent research, however, documents the great degree to which student thinking is dynamic and context-sensitive, implicitly calling for explanations not only of change but also of stability. In other words: When a pattern of student reasoning is sustained in specific moments and settings, what mechanisms contribute to sustaining it? We characterize student understanding and behavior in terms of *multiple local coherences* in that they may be variable yet still exhibit local stabilities. We attribute stability in local conceptual coherences to real-time activities that sustain these coherences. For example, particular conceptual understandings may be stabilized by the linguistic features of a worksheet question, or by feedback from the students' spatial arrangement and orientation. We document a group of university students who engage in multiple local conceptual coherences while thinking about motion during a collaborative learning activity. As the students shift their thinking several times, we describe mechanisms that may contribute to local stability of their reasoning and behavior.

## 1. INTRODUCTION

Research into student knowledge and learning of science has posed conceptual *change* as a primary concern. Research from the 1970s and 1980s tends to characterize broad conditions for and processes of change.[1] Novice misunderstandings are described in terms of singular and often flawed knowledge structures – naïve theories,[2] alternative frameworks,[3] or ontological categories[4]; the events of interest are those in which knowledge structures give way to other structures in a progression toward expertise. More recent work focuses on micro-processes of change, characterizing the diversity of ideas that students employ for making sense of physical phenomena and the nuanced contextuality of these varied ideas across time and setting.[5,6] This later research sees students as gaining expertise by refining and repurposing existing knowledge – phenomenological primitives,[7] conceptual and epistemological resources,[8,9] or even pieces of ontological knowledge.[10]

Both of these perspectives have been primarily concerned with explaining conceptual change, However, research investigating the dynamics of students' thinking over short timescales implicitly calls for explanations of *stability* in student thinking. In the course of interactional processes documented with video of complex learning environments, local patterns of thinking may be seen to stabilize, then destabilize, and then resolve into new patterns. This kind of "change" may only involve dynamics by which the system of intuitive knowledge settles into various states, without implying any change to the system structure itself.

We take the perspective that student understandings reflect *multiple local coherences*.[6,9,11] This specific term is intended to capture the notion that understanding and behavior are often quite variable, yet still exhibit local stabilities. Our guiding question is: When a pattern of student reasoning is sustained in specific moments and settings, what mechanisms contribute to sustaining it? We find that stabilities in student thinking can be attributed to real-time activities that sustain specific understandings. For example, particular conceptual understandings may be sustained by the linguistic features of a worksheet question, or by feedback from the students' spatial arrangement and orientation. We illustrate our approach with an analysis of students' multiple local conceptual coherences during a collaborative learning activity in an introductory college physics classroom.[1] Our perspective contrasts with one in which stability is an inherent feature of knowledge systems.

## 2. INTERPRETING DISTANCE IN A TUTORIAL ON SPEED

Two vignettes below are taken from a student discussion during a collaborative learning activity in introductory college physics. The students are in a *tutorial*, a common instructional format in introductory physics in which students are expected to work through a conceptual worksheet in small collaborative groups.[12] In this particular tutorial (a modified version of a tutorial from Ref. 12), a group of four students in an introductory university physics course discuss "tickertape" representations of motion, used to teach students concepts about velocity and acceleration.

"Tickertape" is a long strip of paper that can be attached to a moving object to record its motion. The record of the motion is produced as the paper is pulled through a tapping device that marks the paper at a constant rate. In this tutorial, strips of tickertape represent the motion of a cart that was recorded prior to class (see Figure 1 for a schematic). Each student has been given a small strip cut from the entire tickertape; the strips all represent the same amount of time (i.e., have the same number of dots) but different speeds (i.e., have different distances between dots).

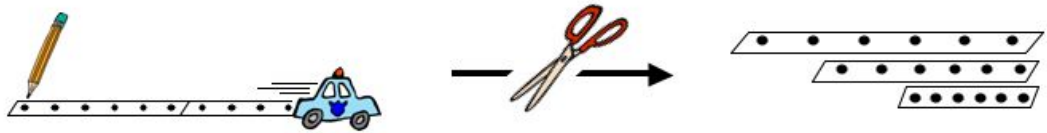

Figure 1: Schematic for how tickertape strips were produced prior to class

Below, the students discuss how the amount of time taken to generate each of their different strips compares, and incorrectly decide (as many groups initially do) that the shorter strips represent less time.

[01:00]

Beth:  Obviously, it takes less time to generate the more closely spaced dots.

John:  So you are saying it takes less time to make the shorter segments?

Beth:  Yeah.

Kate:  How can you tell?

Beth:  You can tell because it's a shorter distance.

Paul:  It's a shorter segment.

[01:36]

---

[1] Some of this material appeared previously in Frank, B. W. *Multiple conceptual coherences in the speed tutorial: Micro-processes of local stability*, a paper presented at the 9th International Conference of the Learning Sciences (ICLS 2010), Chicago, IL.

The students discuss this conclusion for several minutes, pointing out other features of the strips that support this conclusion. They soon move on to the next part of the worksheet, which involves doing some calculations based on distance measurements. After finishing these calculations, the students are prompted by a specific question in their tutorial workbook to reflect on their assumptions in doing these calculations.

[7:45]

Kate: I think we also assumed in that these were made by the speed at which the paper traveled through the tapper, which was different for each paper.

…

John: Right, cause if you move it really fast then *[quick movement of hand]*

Beth: That's true! It could depend on how fast the ribbon was pulled *[moves hand]*

…

Kate: We're assuming that umm-

Beth: That the length is proportional to-

Kate: -The speed at which the ribbon was pulled through.

[8:26]

These two vignettes involve student understandings that are substantively quite different. In the first, students interpret distance as conveying information about time. In the second, students interpret distance as conveying information about speed.

It seems natural, given the change in students' interpretation of what distance means, to ask, "What caused the students to change their thinking?" This question seems especially relevant since they changed their understanding from an incorrect one (thinking that the distance indicates time) to a correct one (that the distance indicates speed). We might consider several plausible accounts of what caused the change: for example, (i) the particular question from the worksheet triggered them to think meta-cognitively about what they had been doing, (ii) engaging in the measurement exercise enabled the students to attend to and coordinate new aspects, and (iii) Kate was privately thinking correctly the entire time and only now found a way to have her ideas introduced and considered. Each of these answers may hold some truth in explaining the change in students' collective understanding.

The question about change, however, may be misleading. Students do not change their thinking on the matter only one time. In particular, it does not seem that the students decide on a particular interpretation of distance (as indicating time, or as indicating speed). Instead, their understanding seems to alternate between two distinct ways of making sense of the strips of paper as representations of motion. Several minutes after the end of the above transcript (11:20), Beth brings up again the idea that the shorter strips take less time. Kate, who initially introduced the idea that the speed causes the different distances, seems to now agree with Beth. As the tutorial goes on, the group regenerates the understanding that distance must be related to speed and not time by attending to acts of pulling and tapping which generated the strips. Then, as the students go back to a previous page to erase their wrong answers, one of the students becomes convinced again of their earlier idea, suggesting that "less time to generate the shorter segments" is still right.

For the researcher, trying to understand the cause of a single moment of change in the students' thinking becomes a problem of explaining multiple changes back and forth. The seemingly straightforward question, "What caused the students to change their thinking?" may not only be difficult to answer; it may be misleading in its premise. For one, there might not be a simple or single causal explanation for any of these transitions. For another, the apparent change in the students' thinking might not reflect change to the structure of students' intuitive knowledge. Instead, the changes we observe might only reflect dynamic transitions among competing understandings, each of which exhibits some local stability. In this perspective, the central problem for researchers is to understand the local stabilities themselves. The aim of this paper is to describe how the students' initial understanding may be stabilized through a variety of different mechanisms.

### 3. THEORETICAL FRAMEWORKS

We take the position that distinct patterns in student thinking reflect dynamic stabilities, and that these stabilities require explanation. The stability of student thinking need not be due to the existence of a robust belief or mental category, but rather may be due to how real-time activity sustains patterns of thought. "Real-time activity" may be understood in a variety of ways with respect to human behavior and learning: for example, in terms of individual, cognitive activity (e.g., connectionism) or in terms of distributed, social activity (e.g., activity theory). We briefly discuss relevant aspects of these knowledge-based and participation-based perspectives.

#### A. Interaction among knowledge agents

Individual cognitive perspectives that emphasize variability describe activity as occurring among fine-grained knowledge structures. For example, the individual mind may be conceptualized as a society of mindless "agents," numerous in kind and in their interactions, which are cognitive elements for perceiving, doing, and remembering.[13] These cognitive elements interact in various ways to generate and represent thinking.[14]

One kind of cognitive element that has been discussed extensively is diSessa's phenomenological primitives[7]—pieces of knowledge reflecting our intuitive sense of mechanism. Thinking about the above vignette from this perspective, we might think of the students' inference about the shorter strips taking less time as resulting from an intuition that *less distance implies less time*. This piece of knowledge for relating space and time is part of how the students interpret and make sense of the strips as representations of motion. Other relevant knowledge pieces might include *going faster implies less time* and *going faster implies more distance*.[15] These knowledge pieces would likely be part of a knowledge system whose parts interact with one another.

diSessa approaches the question of how knowledge pieces may be activated and stabilized in particular contexts in terms of *cuing priorities* and *reliability priorities* that govern their use. Phenomenologically, cuing priorities describe the likelihood that a given piece of knowledge will activate in a given context, and reliability priorities describe the likelihood that a given piece will remain active in a given context. For example, the idea "shorter strips take less time" became active in the above conversation (cuing) and then persisted for several minutes beyond the transcript (reliability). Reliability may also be defined structurally: for knowledge elements in a network, reliability is related to the number of activation pathways leading away from and then back to a given knowledge element. This definition suggests a mechanism for stable activation of an idea in a given setting: other "nearby" ideas support its sustained activation. In considering the vignettes above from this perspective, we might observe whether other ideas arise during the conversation that help to support their continued thinking that the shorter strips take less time, as well as how the context supports the cuing of those ideas.

In the type of cognitive model described above, mechanisms for sustaining patterns of thinking arise from relationships and interactions among knowledge elements. For this reason, we refer to these types of perspectives as "knowledge-based."

B. Interaction among participatory agents

A related group of other perspectives that we will refer to together as "participation-based" describe activity as occurring among persons and artifacts rather than among entities of the mind. The unit of analysis for participation-based perspectives is *persons-in-settings* (rather than individuals or individuals' knowledge).[16,17] Analysis focuses on observable unfolding social activities (rather than knowledge use) and how these activities arise within social settings that are defined by the participants (rather than by outside observers).[18] Material artifacts mediate activity,[19] provide affordances for action,[20] and shape semiotic fields.[21] For example, in the analysis below, a particular physical arrangement of tickertape strips on the table will be shown to stabilize a particular understanding of the motion that the strips represent.

Even as participation-based perspectives prioritize the interactions between people and things, in some cases these perspectives also concern themselves with the substance of the ideas to which participants' activity is oriented. Different ideas are seen as constrained and supported by different patterns of action and social collaboration that take place within changing material settings. Embodied cognitive perspectives emphasize how human understandings have a strong basis in bodily action:[22] for example, the action of pointing might support understanding of objects and their spatial relations, whereas dynamically moving one's hand through space might better support understanding of trajectory or cause and effect relations. Similarly, in social collaboration, establishing joint attention to something in the environment as part of collaborative action might afford different epistemic activities than establishing mutual attention to one another;[23] such social cues help individuals to *frame*[9,24] activities. In participation-based accounts, mechanisms for sustaining patterns of activities arise from the affordances of artifacts and the patterns of participation that take shape around them.

In many respects, these perspectives oppose the above knowledge-based accounts: Lave, for example, in describing the variety of mathematical practices, denies the existence of mental constructs such as arithmetic.[25] Others include both knowledge and participation views in their accounts of students' thinking[26] or have advocated for their commensurability.[9,27] Hutchins, for example, argues that what we think of as "cognitive processes" include not only personal mental operations but also operations on and by the material structure in the physical world.[28,29] We will return to this perspective in the analysis that follows.

4. MECHANISMS FOR SUSTAINING CONCEPTUAL COHERENCES

The goal of this section is to explain the stability of the aforementioned physics students' initial thinking in terms of real-time processes. Both knowledge- and participation-based mechanisms plausibly contribute to the stability of students' thinking.

A. Knowledge-based mechanisms

One general means of accounting for the stability of students' ideas in the kinematics activity is by considering the activity of knowledge elements. These knowledge elements may be influenced by features of the learning environment and by other knowledge elements.

*1. Contextual feedback from setting*

Particular micro-features of the learning context sustain the activation of intuitive knowledge. The specific language of the worksheet question, the salience of length differences among the tickertape strips, and a congruence of part-whole relationships on the strips all contribute to sustaining the idea that *less distance implies less time*.

The question on the worksheet reads, "How does the time taken to generate one of the short segments compare to the time to generate one of the long ones?" The phrasing of the question in terms of two parallel word-sequences that differ only in a single term draws attention to the words "short" and "long," emphasizing distance features of the strips.[21] The words "short" and "long" are used flexibly in everyday language to refer to both distance and time (but no other concepts). Movies may have short duration, roads may go a short distance, and car rides may be short in both senses. The colloquial ambiguity of the word "short" supports the intuition that *less distance* ("shorter") *implies less time* ("shorter"). In response to the worksheet question, Beth states, "Obviously, it takes less time to generate the more closely spaced dots." The use of "short" and "long" (by the worksheet and subsequently by the students) contributes to sustaining the students' initial idea.

The different lengths of the tickertape strips are a highly salient feature, observable even from a distance. The dots, in comparison, are subtly marked. The salience of the different lengths supports students' initial and sustained attention to distance features of the strips, which has been shown to preferentially cue the idea that *less distance implies less time* over other competing ideas when students are asked to made judgments about duration [15]. In the transcript, after Beth gives her initial answer, other students explain that they can tell it takes less time to generate certain strips because those strips are shorter. In these explanations, the students physically indicate distances on the strips with their fingers and pencils. These statements and gestures show that students are closely attending to the lengths of the strips, supporting their initial understanding.

Another feature of the strips that contributes some stability is that the spacing between the dots is longer for strips that are longer overall, and shorter for strips that are shorter overall. (This is because each strip, by design, has the same number of dots.) As students shift their attention from the entire length to the length of the parts (or vice versa), students see the same information. Shorter strips *or* shorter spacings imply a shorter amount of time. Evidence that this shift in attention does not disrupt the students' thinking, and may in fact stabilize it, comes from a part of their conversation that happens shortly after the end of the first transcript above. (Time stamps mark the time since the start of recording, which is approximately the start of the lesson.)

[01:47]

Kate: When we are talking about segments, are we like not thinking about how long the total paper is? Are we just looking at the marks? Are we supposed to be considering

John: I'm guessing they like mean from here to here. *[pointing with pencil between marks]*

Kate: Like I wonder why like the papers are all different lengths.

Beth: Cause none of these papers are the exact same si-ize. Except for these two *[pointing]*

Paul: Right because I think *[moving hand toward the center]* they all have the same amount of dots *[pointing to several locations on one of the strips]*.

Beth: Oh-oh

Paul: I think they all have six dots.

Beth: Oh do, they?

Kate: Is that true? 1, 2, 3, 4, 5, 6 *[pointing to successive dots on a strip]*

Paul: So, it's a shorter amount of time for a shorter piece of paper

[02:17]

Kate asks if they are supposed to be looking at the distance between the "marks" or the total length. As the students coordinate between these two features (realizing that there are six dots on each), the students maintain their understanding that the shorter strips take less time.

2. *Structural feedback from other knowledge*

The students' initial understanding is also stabilized through the recruitment of other ideas that support the initial idea. Two specific ideas arise together with the idea that the shorter strips take less time: *bunched up implies faster,* and *faster implies less time*.

During the student's activity, they attend not only to the length of the entire strip and to the length of its parts, but also to the "density" or proximity of the dots to each other. As they shift their attention to the density of the dots, they start to describe the shorter strips not merely as taking less time but as being faster as well:

[01:37]

Kate: How can we tell? [that it's a shorter amount of time]

Paul: It's a shorter segment

Beth: It's a shorter distance

John: You made more segments in the same amount of space, you've made like more little things

Paul: It's like the same amount of dots in a shorter distance

Beth: Yeah, when given the same length of. [1:48]

*[Discussion of whether length refers to length of paper or distance between dots]*

[2:43] Kate: Wait, what did we say about the six dots?

Beth: It took a different amount of time to generate the dots, six dots for each slip of paper. You can tell because each slip is a different length, but it has the same amount of dots on it. [2:58]

*[Pause while students attend to their worksheets][*

[4:08] Beth:*[reading from worksheet]* How do you know how to arrange [the strips by speed]

Kate: Ahh. The shorter the segments the faster the speed

John: Yeah.

Paul: Also shorter the paper, it's the same thing. [4:30]

The students' initial conclusion – *less distance implies less time* – co-occurs with the another related idea that seems to suggest that *bunched up implies faster*. This idea that *bunched up implies faster* can also be understood to work together with yet another idea that arises: *faster implies less time*. Beth explicitly discusses this idea during their discussion when she says later (11:16), "Wouldn't yours be going slower than mine, because it took more time to make that same?" This statement implies that a certain strip must have gone slower because it took more time. Recalling their earlier conclusion that longer (and less bunched up) strips were slower, a single strip is thus conceived of as being at once either long, widely-spaced, and slow, or short, bunched, and fast. "Faster" and "slower" are another pair of words with ambiguous meanings, in this case potentially referring to either frequency or speed: a clock may tick faster, a car may travel faster, and a turning crank may increase both its frequency and its speed simultaneously. The words "short" and "fast" have multiple meanings that overlap, supporting the association of *short, bunched,* and *fast* with one another.

These three elements of intuitive knowledge–*less distance implies less time, bunched up means faster*, and *faster implies less time*– are all connected through their linguistic and conceptual overlap. Ideas of *less distance* and *less time* are bound by a linguistic overlap with the words "longer" and "shorter". *Less distance* and *bunched* are bound by their conceptual overlap with a spatial sense of proximity. *Bunched* and *less time* (and even notions of speed) are bound by their linguistic overlap with the word "faster" and "slower." This network of ideas, illustrated in Figure 2, can be understood to exhibit stability through the mutual relationships among its parts.

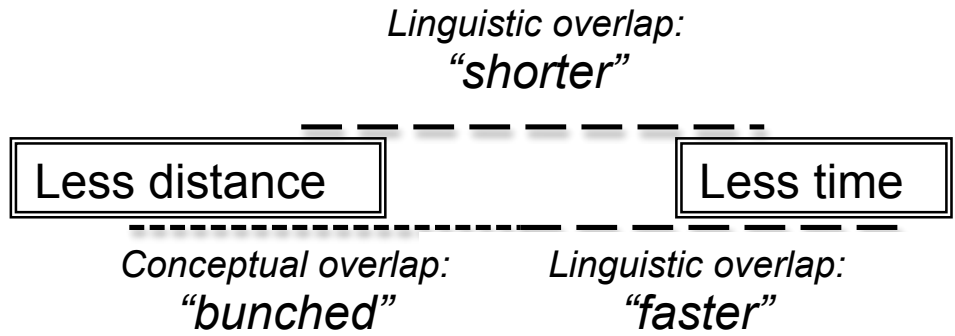

Figure 2. Connections among intuitive knowledge elements in the tickertape scenario. Long dashes signify linguistic overlap; short dashes signify conceptual overlap.

These knowledge-based mechanisms describe how relationships among specific elements of knowledge and contextual features provide some stability to the students' understanding, contributing to its persistence over time.

### B. Participation-based mechanisms

So far, we have described possible mechanisms for stabilizing student thinking that involve contextual features and knowledge elements. In this section, we suggest mechanisms involving interactions among the students and with artifacts.

#### 1. Feedback from participants' spatial arrangement and orientation

Students arrange themselves spatially in distinctive ways during the tutorial, both relative to each other and relative to artifacts around them. At the beginning of the tutorial (and when the students initially agree that the longer strips take more time), the students collectively participate in two particular *behavioral clusters* – sets of behaviors that largely occur together among all the individuals, with some exceptions.[9] The two distinct clusters are (i) behaviors oriented toward their worksheets (Figure 3) – a pattern described by Scherr and Hammer,[9] and (ii) behaviors oriented toward the strips at the center of the table (Figure 4). As in Scherr and Hammer,[9] peers are observed to participate primarily in the same behavioral cluster. Though individual students opt in and out of each of these clusters, the collective structure of participation persists over several minutes at a time. These spatial arrangements establish and stabilize students' joint attention, contributing to the durability of specific intellectual activities.

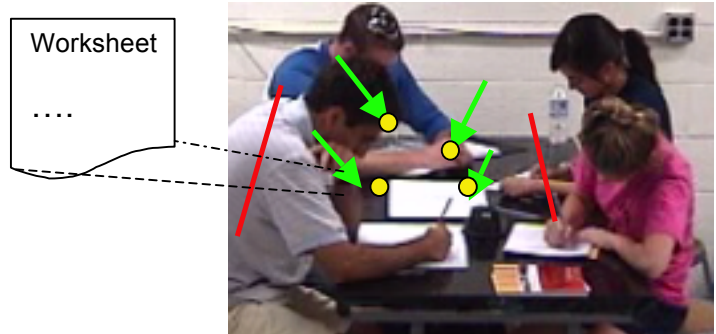

Figure 3. Student behaviors oriented toward the worksheets

The tickertape-oriented behaviors include directing gaze, gesture, and bodies toward the strips at the center of the table. Students lean in, look inward, and position their hands at the center of the table. The students' behavioral cluster is dynamically stable in the sense that any one student's behavior is coupled to the behaviors of the others. As students lean toward the center and point to strips and locations, other students look in that direction, leaning in and pointing to the strips as well. At the same time, students notice, describe, and compare features of the strips (e.g., *short distance means short times* and *bunched up dots means fast*). Joint maintenance of a spatial-orientational system and joint attention to community objects couples to the substance of the ideas that students discuss.

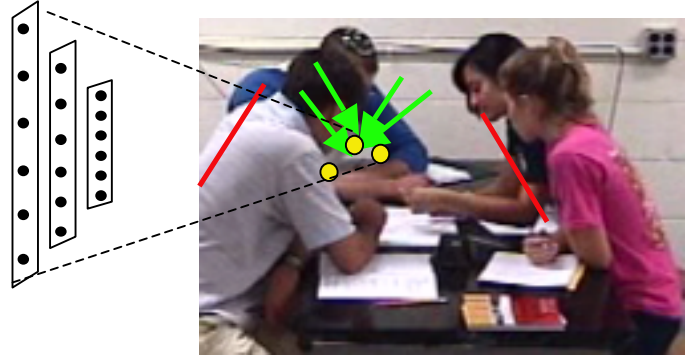

Figure 4. Student behaviors oriented toward the strips

The worksheet-oriented behaviors, in contrast, include hunching over, looking down, and using hands to write and point to worksheet locations. While participating in this behavioral cluster, the students neither discuss aspects of the physical motion of the strips, nor attend to one another. Instead, they read questions out loud from their worksheets, suggest brief answers to each other, and write in their worksheets.

During the initial discussion in which students discuss how the longer strips take more time, the students spend nearly equal time exhibiting behaviors within the two clusters described above. In total, the students transition eight times as they jointly make comparisons of the strips and establish answer to write down.

The two behavioral clusters described above contrast with a third cluster in which students orient to each other: sitting upright, looking to each other, and gesturing in animated ways (Figure 5). In this third behavioral cluster, students begin to discuss ideas about how the strips were made, discussing and even simulating how the paper moved and how the tapping device created the dots. (In the tickertape-oriented behavioral cluster, they discussed static features of the strips.) While discussing these ideas, the students conclude that the longer strips represent faster motion, which is different than what they concluded in other parts of the discussion.

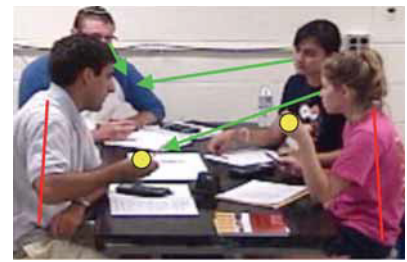

Figure 5. Student behaviors oriented toward each other

Students move in and out of these behavioral patterns with remarkable synchrony. For example, when they change their orientation from the worksheets to the strips, they do so in a sudden cascade: one student leans into the strips, then another student looks over and leans in, and then all the students are leaning in examining the strips, all within a couple seconds of the first transition. In another example, Kate lifts a strip of paper off the table and presents it to the group: all of the students simultaneously look at the strip of paper and begin discussing how it was made. At some point in the discussion, Kate removes the strip of paper from view, hiding it behind her shoulder, and the students cease their discussion, shifting their gaze back toward their worksheets. The transitions are precisely coordinated (though not centrally directed), suggesting that feedback from participants' spatial arrangement and orientation helps to stabilize local patterns of thinking.

### 2. *Material affordances and constraints*

The three behavioral clusters above, which help to characterize the students' activity, each take place in a particular material setting that provides affordances for these behaviors. In particular, the behavioral cluster involving joint attention to the tickertape strips is supported by the strips being located in the center of the table; the behavioral cluster involving individual focus on worksheets is supported by the fact that there is a worksheet in front of each student; and the third behavioral cluster, of mutual attention and discussion, is supported by one student holding a strip of tickertape in her gesture space and using it as part of a simulation of the motion it represents. These three specific material environments establish and stabilize students' joint attention much as did the spatial arrangement of their bodies, helping to sustain specific intellectual activities. The three environments are in part constructed by the students: they are the ones to place the strips.

The arrangement of strips side-by-side in the center of the table stabilizes reasoning that relies on comparing features of the strips, such as *less distance implies less time* and *bunched up means faster*. "Stacking" the strips as in Figure 1 provides a "material anchor" for comparing the lengths of the strips, reducing the cognitive effort in making the comparison and promoting reasoning that relies on that comparison (*i.e.,* ordering time and speed by means of length). [28] In addition, the "stacked" arrangement supports certain forms of gesture, especially pointing with a single finger within the area of joint attention, or indicating a space with two fingers. These embodied actions may support students' ideas about spatial relations among parts of objects, such as the number and density of dots.[30]

Later, students take possession of individual strips, and engage in reasoning that can be seen as coordinated with the change in the material setting. When Kate lifts a single strip off the table, she is in a position to physically enact the pulling of the strips by the cart; she does so as she explains how she thinks the strip was made. When she does so, the other students in the group look at her (instead of at the stacked strips), engaging in a new pattern of mutual attention as they discuss aspects of the physical motion that made the strips. Their discussion is concerned with imagined events of which the individual strips are the durable traces,[17] rather than comparisons of features of the strips. A change to their material setting supported different options for embodied action and participation.

Including the material setting as one of the influences on student thinking creates a theoretical account in which the stability of students' thinking is coupled to the stability of the physical world. The strips, once stacked at the center of the table, remain there until they are rearranged. The worksheets, similarly, stay in their locations, one in front of each student. This stable material arrangement makes it natural for students to coordinate behaviors oriented to the stacked strips with behaviors oriented to the worksheet. Their careful attention to just the visual properties of the strips (*i.e.,* not to any sense of motion) in order to answer specific worksheet questions takes place in a setting that supports its persistence. The students seamlessly move their attention back and forth between the stacked strips and the worksheets repeatedly during the first ten minutes of tutorial, sustaining not only particular patterns of attention and action upon the strips, but also the use of specific intuitive knowledge. Their attention, action, and knowledge are stabilized in part by the physical stability of the material artifacts with which they interact. Participation-based perspectives that are oriented to the role of "things" in cognition may pose these material artifacts as cognitive agents, whose functions in the activity may include remembering important values of quantities, prompting elaborate conceptual frameworks, or drawing attention to specific features.[29,31]

## 5. DISCUSSION

The primary goal of this paper is theory development. Specifically, we hope to have identified a variety of possible mechanisms for a certain learning phenomenon: the sustaining of a one possible understanding for several minutes in a collaborative learning activity. The phenomenon is one that, in some education research paradigms, would be considered more as an absence of a phenomenon in that it is an absence of conceptual change. We hope that part of the contribution of our work is to position the stabilization of reasoning patterns as a type of occurrence worthy of analysis.

In the service of this larger goal we characterize the stability of a single group of students' initial thinking during a collaborative learning activity in terms of a variety of micro-processes that can help to sustain a reasoning pattern for an extended time. We explain the local persistence of their understanding, rather than the causes of change. We explain this stability both in terms of how elements of students' intuitive knowledge are reliably activated in the setting and in terms of how students' interactional behaviors dynamically constitute and stabilize aspects of this setting. Whether these various knowledge-based and participation-based mechanisms act together to jointly influence student reasoning, or whether they are alternative theoretical accounts, is not answered by this analysis.

We claim that the students make their initial inference about the tickertapes based on a shared intuition that *less distance implies less time*. There is evidence that each of the students makes sense of the situation using this idea at various times, and often times they do so together. Particular features of the context (specific features of the worksheet questions and of the strips themselves) contribute to activation of that knowledge. Other knowledge recruited to make sense of the situation provides a stable network of ideas for interpreting the patterns they notice. This network of ideas arises within a social context of discussing patterns and writing down answers to worksheet questions. During their initial discussion, the students orient their gazes, gestures, and bodies toward the collection of the strips at the center of the table and toward the worksheets in front of them; they do not engage in behaviors of mutual attention until much later. This activity of noticing and describing patterns to be written in their worksheets is both stabilized by the students' specific (collective) behaviors in this activity and by the location and arrangement of material artifacts that are central to the activity.

This characterization, while focused on explaining the stability of the students' initial pattern of thinking, also suggests an account of the dynamics of their subsequent understanding: their understanding changes when the constraints and affordances of their activity change. The students' thinking co-evolves with changes to their patterns of intellectual attention (to physical mechanisms involved in the motion), changes to the location of material objects (individual strips are held by individual students instead of collected in the center of the table), and to the patterns of interactional behavior (students look at each other as they articulate complex ideas). These changes are new patterns that take hold with their own local stability. A goal of further research is to begin to understand which aspects of activity are ephemeral and which continue to exert influence.

Beyond this particular case, considering students' intuitive thinking as reflecting multiple conceptual coherences may be useful for describing stabilities of thinking and behavior that persist on longer time scales (e.g., Why do students appear to hold on to robust misconceptions?). In the long term, we hope to both account for the variability of student understandings and also describe how, in particular contexts, understandings can settle into patterns indicating common and robust ideas. This program might be achieved through careful attention to the multitude of cognitive and social mechanisms that contribute to the local stability of reasoning as it occurs within activity.

ACKNOWLEDGMENTS

The research described here was conducted at the University of Maryland. The authors would like to thank David Hammer, Joe Redish, Andy Elby, Ayush Gupta, Luke Conlin, Renee Michelle Goertzen, Michael C. Wittmann, and Amy D. Robertson for their helpful conversations. The research has been funded in part by the National Science Foundation under Grant Nos. REC-0440113, REC-0633951, and DRL 0822342. Any opinions, conclusions, or recommendations expressed in this material are those of the author and do not necessarily reflect the views of the National Science Foundation.